\documentstyle[12pt,epsf]{article}

\begin{document}

\baselineskip=18.8pt plus 0.2pt minus 0.1pt

\makeatletter

\renewcommand{\thefootnote}{\fnsymbol{footnote}}
\newcommand{\beq}{\begin{equation}}
\newcommand{\eeq}{\end{equation}}
\newcommand{\bea}{\begin{eqnarray}}
\newcommand{\eea}{\end{eqnarray}}
\newcommand{\nn}{\nonumber}
\newcommand{\hs}[1]{\hspace{#1}}
\newcommand{\vs}[1]{\vspace{#1}}
\newcommand{\Half}{\frac{1}{2}}
\newcommand{\p}{\partial}
\newcommand{\ol}{\overline}
\newcommand{\wt}[1]{\widetilde{#1}}
\newcommand{\ap}{\alpha'}
\newcommand{\bra}[1]{\left\langle  #1 \right\vert }
\newcommand{\ket}[1]{\left\vert #1 \right\rangle }
\newcommand{\vev}[1]{\left\langle  #1 \right\rangle }
\newcommand{\bm}[1]{\mbox{\boldmath $ #1 $}}

\makeatother

\begin{titlepage}
\title{
\hfill\parbox{4cm}
{\normalsize KUNS-1622\\{\tt hep-th/9912029}}\\
\vspace{1cm}
Non-BPS D0-brane instanton effects in type I string theory
}
\author{Yoji Michishita
\thanks{
{\tt michishi@gauge.scphys.kyoto-u.ac.jp}
}
\\[7pt]
{\it Department of Physics, Kyoto University, Kyoto 606-8502, Japan}
}

\date{\normalsize December 4, 1999}
\maketitle
\thispagestyle{empty}

\begin{abstract}
\normalsize
 
We investigate the instanton effects of non-BPS D0-brane in type I
string theory. We argue the general properties of these instanton
effects and consider these on $R^9\times S^1$ as a simple example
using the effective action of D0-brane.

\end{abstract}
\end{titlepage}

\section{Introduction}

D-branes give us various nonperturbative informations of string
theories. One of them is the D-brane instanton effect \cite{bbs}.
Euclidean D-branes wound around nontrivial cycles which are the
solutions of equation of motion of worldvolume action contribute
largely to the path integral (or its counterpart of string theory),
and give corrections to amplitudes or
low energy effective action. The way to calculate instanton effects by
using effective action is discussed in \cite{bbs,hm}
and stringy calculation is presented in \cite{gg}.
So far BPS D-branes are used to calculate instanton effects. However,
various non-BPS stable D-branes are discovered recently. 
(For reviews see \cite{rev}.)
Since these
are stable they give large contributions to the path integral
similarly to BPS D-branes.

In this paper we argue non-BPS D0-brane instanton effects in type I
string theory as an example of such effects. Non-BPS D0-brane has
$SO(32)$ spinor charge and is a counterpart of perturbative massive 
$SO(32)$ spinor states in heterotic $SO(32)$ theory \cite{s1}.
This brane forms a long
multiplet of supersymmetry and therefore is non-BPS. However, $SO(32)$
spinor charge protect it from decay. Its tension is $\sqrt{2}$ times
larger than that of BPS D-branes in the weak coupling regime, and
is modified as the coupling constant grows. Therefore instanton action
is a complicated function of dilaton. Since we do not know this
function we restrict ourselves to the weak coupling regime.

Non-BPS D-branes have more fermion massless modes living on them than
BPS D-branes and we need more vertex operators to saturate their zero
modes as will be seen in a concrete example. Hence non-BPS D-brane
instanton corrections are higher order effects than that of BPS
D-branes.

This paper is organized as follows.
In section 2, we briefly review the non-BPS D0-brane and discuss
general properties of its instanton effects. In section 3, we review
the superfield formalism of type I string theory and construct the
effective action and fermion vertex operators following \cite{s3}.
In section 4, we consider $S^1$ compactification as a simple example
and calculate some simple amplitudes by using the effective action
constructed in section 3. Section 5 contains some discussions.

\section{Non-BPS D0-brane instanton}

Non-BPS stable D0-brane in type I string theory can be constructed
from D1-$\ol{{\rm D1}}$ system through the tachyon condensation
mechanism \cite{s1}
and has no RR charge. Therefore its boundary state has NSNS part
only and the overall factor is $\sqrt{2}$ times that of BPS D-brane
\cite{s2,fgls}:
\beq
\ket{D0}=\sqrt{2}\ket{{\rm NS}+{\rm NS}+}.
\eeq
This factor $\sqrt{2}$ means that the tension of non-BPS D0-brane is
$\sqrt{2}$ times larger than that of BPS D-brane:
\beq
 T_{D0}=\Half\cdot\sqrt{2}\cdot\frac{1}{\sqrt{\ap}g_s}.
\eeq
Here the factor $\Half$ comes from the orientifold projection and 
$g_s$ is the string coupling constant. This value is valid in the
weak string coupling regime and is corrected as the coupling constant
grows, unlike BPS D-branes. In the strong coupling regime the
D0-brane corresponds to the perturvative massive states with $SO(32)$
spinor charge of dual heterotic theory. Their tension (mass) is
$\frac{2}{\sqrt{\ap_{het}}}=\sqrt{\frac{2}{\ap g_s}}$, where the right
hand side is written in terms of the quantities of Type I theory.
Thus the exact tension is expressed in terms of an unknown function
$f(g_s)$:
\bea
 & T_{D0}=\frac{1}{\sqrt{2\ap}}f(g_s),& \nn\\
 & f(g_s)\rightarrow \frac{1}{g_s}
 \quad{\rm as}\quad g_s\rightarrow 0, \quad
 f(g_s)\rightarrow \frac{2}{\sqrt{g_s}}\quad{\rm as}\quad
 g_s\rightarrow \infty. &
\eea
Since we do not know what $f(g_s)$ is, we restrict ourselves to
the weak coupling regime.

The massless modes living on the D0-brane are scalars $X^m$
corresponding to ten spacetime directions, one Majorana-Weyl spinor
$\theta^\alpha$ as a superpartner of $X^m$, and a fermion $\xi^I$ with 
a $SO(32)$ vector index $I$ \cite{s2}.
$X^m$ and $\theta^\alpha$ come from D0-D0 strings and $\xi^I$
from D0-D9 strings. Quantization of zero modes of $\xi^I$ gives
$SO(32)$ spinor charge.

Non-BPS D0-brane is stable only when it is alone.\footnote{There are
some exceptions when some spacetime directions are compactified.
For example, when two D0-branes are placed on the antipodal points of
$S^1$ respectively, this configuration is stable.}
When there are two 
D0-branes, they attract each other since there is no RR
repulsive force and are pair-annihilated. This can be seen from the
fact that the corresponding K-theory group is ${\bf Z}_2$ \cite{w}.
The presence of branes of other type also destabilizes the
configuration for the same reason as above. (NSNS force derived from
cylinder amplitude vanishes only when the number of directions with
Neumann boundary condition on the one end and Dirichlet on the other
is four. (See e.g. \cite{bvflprs}.) But there are no branes which can
make such configurations with a D0-brane in type I string theory.)

If a direction transverse to a D0-brane is compactified with the
radius $R$, it decays to a D1-$\ol{{\rm D1}}$ system for
$R<\sqrt{\frac{\ap}{2}}$. This can be understood by keeping track of
the construction given in \cite{s1}
and also by the following argument.

If there is some tachyon mode on a brane, then it is unstable. We can
see whether there are any tachyon mode or not by computing the
cylinder and M\"{o}bius amplitude. Their sum ${\cal A}$ is
\bea
{\cal A} & = & V\int_0^\infty\frac{dt}{2t}(8\pi^2\ap t)^{-\Half}
\biggl(\sum_{w\in {\bf Z}}e^{-\frac{2\pi}{\ap}R^2w^2t}
\biggl[\frac{f_3(e^{-\pi t})^8}{f_1(e^{-\pi t})^8}
-\frac{f_2(e^{-\pi t})^8}{f_1(e^{-\pi t})^8}\biggr]
\nn\\
& & +16e^{-i\frac{\pi}{4}}\frac{f_3(ie^{-\pi t})^9 f_1(ie^{-\pi t})}
 {f_2(ie^{-\pi t})^9f_4(ie^{-\pi t})}
-16e^{i\frac{\pi}{4}}\frac{f_4(ie^{-\pi t})^9 f_1(ie^{-\pi t})}
 {f_2(ie^{-\pi t})^9f_3(ie^{-\pi t})} \biggr)
\nn\\
& = & V\int_0^\infty\frac{dt}{2t}(8\pi^2\ap t)^{-\Half}
\biggl(\sum_{w\neq 0}e^{-\frac{2\pi}{\ap}R^2w^2t}
[e^{\pi t}-8+O(e^{-\pi t})]+2+O(e^{-\pi t})\biggr) , \label{cmamp}
\eea
where
\bea
& f_1(q)=q^{\frac{1}{12}}\prod_{n=1}^\infty(1-q^{2n}),\quad
f_2(q)=\sqrt{2}q^{\frac{1}{12}}\prod_{n=1}^\infty(1+q^{2n}), & \nn\\
& f_3(q)=q^{-\frac{1}{24}}\prod_{n=1}^\infty(1+q^{2n-1}),\quad
f_4(q)=q^{-\frac{1}{24}}\prod_{n=1}^\infty(1-q^{2n-1}). &
\eea
If the exponent of $e^{-\frac{2\pi}{\ap}R^2w^2t+\pi t}$ in eq.
(\ref{cmamp}) is positive,
then this factor is divergent as $t\rightarrow\infty$ and represents
the contribution of tachyon. Therefore for $R<\sqrt{\frac{\ap}{2}}$
the D0-brane is unstable.

Since D0-brane is stable for sufficiently large transverse radius, it
gives large contribution to the path
integral as a kind of instanton backgrounds. In order to give finite
contribution, it must be Euclidean and wrapped around some
compactified direction. For example, a D0-brane wrapped once around
$S^1$ with radius $R$ gives the instanton action $\exp(-T_{D0}\cdot
2\pi R)=\exp(-\frac{2\pi R}{\sqrt{2\ap}g_s})$ at the weak coupling
regime. The sum of its
cylinder and M\"{o}bius amplitude ${\cal A}'$ is
\bea
{\cal A}' & = & \int_0^\infty\frac{dt}{2t}\sum_{n\in {\bf Z}}
e^{-\frac{t}{T}(\frac{n}{R})^2}
\biggl[\frac{f_3(e^{-\pi t})^8}{f_1(e^{-\pi t})^8}
-\frac{f_2(e^{-\pi t})^8}{f_1(e^{-\pi t})^8}
\nn\\
& & +16e^{-i\frac{\pi}{4}}\frac{f_3(ie^{-\pi t})^9 f_1(ie^{-\pi t})}
 {f_2(ie^{-\pi t})^9f_4(ie^{-\pi t})}
-16e^{i\frac{\pi}{4}}\frac{f_4(ie^{-\pi t})^9 f_1(ie^{-\pi t})}
 {f_2(ie^{-\pi t})^9f_3(ie^{-\pi t})} \biggr]
\nn\\
& = & \int_0^\infty\frac{dt}{2t}\sum_{n\in {\bf Z}}
e^{-\frac{t}{T}(\frac{n}{R})^2}
[2+O(e^{-\pi t})] ,
\eea
This shows that there is no tachyon mode and the winding D0-brane
is stable. As for multi-instanton configurations, though these are
unstable as explained above\footnote{Except for the cases mentioned in
the previous footnote.}, the D0-branes are infinitely massive in the
weak coupling regime and stable approximately. Therefore these
configurations can give large contributions to the path integral in
the weak coupling regime.

D0-brane instanton effects give corrections to the low energy
effective action of type I string theory. D0-brane has more fermion
massless modes than BPS branes and therefore more fermion zero modes.
We have to insert more vertex operators in order to soak up these zero
modes. Hence its leading contribution appears in higher
derivative terms in the bulk effective action than that of BPS branes.
In type I string theory the
BPS D-brane with the lowest worldvolume dimension is D1-brane and its
instanton effects appear when at least two dimensions are
compactified. But the D0-brane instanton effects emerge even when only 
one dimension is compactified. In the following sections we use the
effective action of D0-brane to calculate the instanton effects.
\section{The effective action and vertex operators 
for non-BPS D0-brane}
In this section we briefly review the superfield formalism of type I
string theory and construct the effective action and vertex operators
of D0-brane following \cite{s3}.
The results of this section will be used in the next section. We use
the notation and results of \cite{adr}.

To describe type I string theory we need supervielbein
$\bm{E}_M^{\;\;A}$, superconnection
$\bm{\Omega}_{MA}^{\;\;\;\;\;\;B}$,
RR 2-form $\bm{B}_{MN}$, and $SO(32)$ gauge field $\bm{A}_M$. 
Supertorsion $\bm{T}$, supercurvature $\bm{R}$, and field strengths
$\bm{H}\equiv d\bm{B}$ and $\bm{F}\equiv d\bm{A}$ satisfy the
Bianchi identities:
\bea
D\bm{T}^A & = & \bm{E}^B\bm{R}_B^{\;\; A}, \\
D\bm{R}_A^{\;\; B} & = & 0, \\
d\bm{H} & = & c_1{\rm tr}\bm{F}^2, \\
D\bm{F} & = & 0.
\eea
Here $c_1$ is a certain constant. We impose the following
constraints\cite{adr}. 
\bea
\bm{T}_{\alpha\beta}^{\;\;\;\;a}=2(\Gamma^a)_{\alpha\beta}, & & 
\bm{T}_{\alpha a}^{\;\;\;\;b}=0, \nn\\
\bm{T}_{a\alpha}^{\;\;\;\;\beta}
 =(\Gamma_a\bm{\Psi})_\alpha^{\;\;\beta},& & 
\bm{T}_{\alpha\beta}^{\;\;\;\;\gamma}=0,\\
\bm{H}_{\alpha\beta\gamma} & = & 0, \\
\bm{F}_{\alpha\beta} & = & 0,
\eea
where $\bm{\Psi}^{\alpha\beta}$ is defined by the above equation. The
solutions of the Bianchi identities are \cite{adr}
\bea
\bm{\Psi}^{\alpha\beta} & = &
 -\frac{1}{24}\bm{T}_{abc}(\Gamma^{abc})^{\alpha\beta}, \\
\bm{H}_{abc} & = & -\frac{3}{2}\bm{\Phi}\bm{T}_{abc}
 +\frac{c_1}{4}(\Gamma_{abc})_{\alpha\beta}
 {\rm tr}[\bm{\chi}^\alpha\bm{\chi}^\beta], \\
\bm{H}_{ab\alpha} & = &
 -\frac{1}{2}(\Gamma_{ab})_\alpha^{\;\;\beta}\bm{\lambda}_\beta, \\ 
\bm{H}_{a\alpha\beta} & = & \bm{\Phi}(\Gamma_a)_{\alpha\beta}, 
\label{phi} \\
\bm{F}_{a\alpha} & = & (\Gamma_a)_{\alpha\beta}\bm{\chi}^\beta, 
\label{chi}
\eea
where $\bm{\Phi}$ and $\bm{\chi}^\alpha$ are defined by equations
(\ref{phi}) and (\ref{chi}) respectively, and
$\bm{\lambda}_\alpha\equiv D_\alpha\bm{\Phi}$.
We can set the components of $\bm{E}_M\,^A$,
$\bm{\Omega}_{MA}\;^B$, $\bm{B}_{MN}$ and $\bm{A}_M$ as follows
using the local symmetries:  
\beq
\bm{E}_m^{\;\;a}|_{\theta=0}\equiv e_m^{\;\;a},\quad
\bm{E}_m^{\;\;\alpha}|_{\theta=0}\equiv\psi_m^{\;\;\alpha},\quad
\bm{E}_\mu^{\;\;a}|_{\theta=0}=0,\quad
\bm{E}_\mu^{\;\;\alpha}|_{\theta=0}=\delta_\mu^{\;\;\alpha},
\eeq
\beq
\bm{\Omega}_{ma}^{\;\;\;\;\;b}|_{\theta=0}\equiv
\omega_{ma}^{\;\;\;\;\;b},\quad
\bm{\Omega}_{\mu a}^{\;\;\;\;b}|_{\theta=0}=0,
\eeq
\beq
\bm{B}_{mn}|_{\theta=0}\equiv B_{mn},\quad
\bm{B}_{m\nu}|_{\theta=0}=0,\quad
\bm{B}_{\mu\nu}|_{\theta=0}=0,
\eeq
\beq
\bm{A}_m|_{\theta=0}\equiv A_m,\quad
\bm{A}_\mu|_{\theta=0}=0,
\eeq
and we set
\beq
\bm{T}_{abc}|_{\theta=0}\equiv T_{abc},\quad
\bm{\Phi}|_{\theta=0}\equiv\Phi,\quad
\bm{\lambda}_\alpha|_{\theta=0}\equiv\lambda_\alpha,\quad
\bm{\chi}^\alpha|_{\theta=0}\equiv\chi^\alpha .
\eeq
We can determine the lower components of  $\bm{E}_M^{\;\;A}$,
$\bm{\Omega}_{MA}^{\;\;\;\;\;\;B}$, $\bm{B}_{MN}$, $\bm{A}_M$ and
$\bm{\Phi}$ by comparing the solutions of Bianchi identities and the
definitions of $\bm{T}$, $\bm{R}$, $\bm{H}$ and $\bm{F}$ in the
appropriate gauge fixing conditions:
\bea
\bm{E}_m^{\;\;a} & = & e_m^{\;\;a}-2\theta^\alpha
(\Gamma^a)_{\alpha\beta}\psi_m^{\;\;\beta}
\nn\\
& & +\frac{1}{4}\theta^\alpha(\Gamma^{abc})_{\alpha\beta}\theta^\beta
\omega_{mbc}
-\frac{1}{24}\theta^\alpha(\Gamma^{bcd})_{\alpha\beta}\theta^\beta
e_m^{\;\; a}T_{bcd}
\nn\\
& & +\frac{1}{8}\theta^\alpha(\Gamma^{abc})_{\alpha\beta}\theta^\beta
e_m^{\;\; d}T_{bcd}
+\frac{1}{8}\theta^\alpha(\Gamma_m^{\;\;\;bc})_{\alpha\beta}
\theta^\beta T^a_{\;\; bc}+O(\theta^3),
\label{compe1}\\
\bm{E}_m^{\;\;\alpha} & = & \psi_m^{\;\;\alpha}
+\theta^\beta\omega_{m\beta}^{\;\;\;\;\;\alpha}
+\frac{1}{24}\theta^\beta(\Gamma_m\Gamma^{abc})_\beta^{\;\;\alpha}
T_{abc}
+O(\theta^2),
\label{compe2}\\
\bm{E}_\mu^{\;\;a} & = & (\Gamma^a)_{\mu\alpha}\theta^\alpha
+ O(\theta^2),
\label{compe3}\\
\bm{E}_\mu^{\;\;\alpha} & = & \delta_\mu^{\;\;\alpha} + O(\theta^2),
\label{compe4}
\eea
\bea
\bm{\Omega}_{ma}^{\;\;\;\;\;b} & = & \omega_{ma}^{\;\;\;\;\;b}
+O(\theta), \\
\bm{\Omega}_{\mu a}^{\;\;\;\;\;b} & = &
\frac{1}{12}(\Gamma_a^{\;\;bcde})_{\mu\alpha}\theta^\alpha T_{cde}
+\frac{3}{2}(\Gamma^c)_{\mu\alpha}\theta^\alpha T_{ca}^{\;\;\;\;b}
+O(\theta^2),
\eea
\bea
\bm{B}_{mn} & = &
B_{mn}-\Half\theta^\alpha(\Gamma_{mn})_\alpha^{\;\;\beta} 
\lambda_\beta
\nn\\
& & +2\theta^\alpha(\Gamma_{[m})_{\alpha\beta}
\psi_{n]}^{\;\;\beta}\Phi
+2c_1\theta^\alpha(\Gamma_{[m})_{\alpha\beta}
{\rm tr}[A_{n]}\chi^\beta] 
+ O(\theta^2),
\\
\bm{B}_{m\nu} & = & -\Half(\Gamma_m)_{\nu\alpha}\theta^\alpha\Phi
+ O(\theta^2),
\\
\bm{B}_{\mu\nu} & = & 0 + O(\theta^2),
\eea
\bea
\bm{A}_m & = & A_m -\theta^\alpha(\Gamma_m)_{\alpha\beta}\chi^\beta
+ O(\theta^2),
\label{compa1}\\
\bm{A}_\mu & = & 0+ O(\theta^2),
\label{compa2}
\eea
\bea
\bm{\Phi} & = & \Phi+\theta^\alpha\lambda_\alpha
+\frac{1}{12}\theta^\alpha(\Gamma^{abc})_{\alpha\beta}\theta^\beta
\Phi\Bigl(T_{abc}+\frac{3}{8}c_1(\Gamma_{abc})_{\gamma\delta}
{\rm tr}[\chi^\gamma\chi^\delta]\Bigr)+ O(\theta^3),
\label{compp}
\eea
where
\bea
T_{abc} & = & -\frac{2}{3}\Phi^{-1}\wt{H}_{abc}
+2\psi_{[a}^{\;\;\alpha}(\Gamma_b)_{\alpha\beta}\psi_{c]}^{\;\;\beta} 
\nn\\
& & -\psi_{[a}^{\;\;\alpha}(\Gamma_{bc]})_\alpha^{\;\;\beta}
\lambda_\beta\Phi^{-1}
+\frac{1}{6}c_1(\Gamma_{abc})_{\alpha\beta}
{\rm tr}[\chi^\alpha\chi^\beta] 
\Phi^{-1},
\\
\wt{H}_{abc} & = & 3\p_{[m}B_{np]}
+c_1{\rm tr}[6A_{[m}\p_nA_{p]}-4A_{[m}A_nA_{p]}],
\\
\omega_{abc} & \equiv & e_a^{\;\;m}\omega_{mbc}
\nn\\
& = & \omega^{(0)}_{abc}+\Half T_{abc}
-\psi_a^{\;\;\alpha}(\Gamma_c)_{\alpha\beta}\psi_b^{\;\;\beta}
-\psi_c^{\;\;\alpha}(\Gamma_b)_{\alpha\beta}\psi_a^{\;\;\beta}
+\psi_b^{\;\;\alpha}(\Gamma_a)_{\alpha\beta}\psi_c^{\;\;\beta},
\\
\omega^{(0)}_{abc} & = &
e_a^{\;\;n}e_b^{\;\;m}\p_{[m}e_{n]c}+e_c^{\;\;n}e_a^{\;\;m}
\p_{[m}e_{n]b} 
-e_b^{\;\;n}e_c^{\;\;m}\p_{[m}e_{n]a}.
\eea
On the flat background we can determine all the components:
\beq
\bm{E}_m^{\;\;a}= \delta_m^{\;\;a},\quad
\bm{E}_m^{\;\;\alpha}=0,\quad
\bm{E}_\mu^{\;\;a}=(\Gamma^a)_{\mu\alpha}\theta^\alpha,\quad
\bm{E}_\mu^{\;\;\alpha}=\delta_\mu^{\;\;\alpha},
\label{flate}
\eeq
\beq
\bm{\Omega}_{MA}^{\;\;\;\;\;\;B}=0,
\label{flato}
\eeq
\beq
\bm{B}_{mn}=0,\quad
\bm{B}_{m\nu}=-\Half(\Gamma_m)_{\nu\alpha}\theta^\alpha\Phi,\quad
\bm{B}_{\mu\nu}=0,
\label{flatb}
\eeq
\beq
\bm{A}_M=0,
\label{flata}
\eeq
\beq
\bm{\Phi}=\Phi={\rm constant},
\label{flatp}
\eeq
\beq
\bm{T}_{abc}=0.
\label{flatt}
\eeq
To relate $e_m^{\;\;a}$ to the string metric we need the following
field (re)definition.
\bea
\Phi & = & e^{-\frac{2}{3}\phi}, \\
e_m^{\;\;a} & \rightarrow & e^{-\frac{1}{6}\phi}e_m^{\;\;a},
\eea
where $\phi$ is the dilaton.\footnote{To relate $\psi_m^{\;\;\alpha}$, 
$\lambda_\alpha$ and $\chi^\alpha$ to the standard gravitino,
dilatino and gaugino respectively, we need the redefinition something
like $\psi_m^{\;\;\alpha}\rightarrow a_1e^{a_2\phi}\psi_m^{\;\;\alpha}
+a_3e^{a_4\phi}(\Gamma_m)^{\alpha\beta}\lambda_\beta$,
$\lambda_\alpha\rightarrow a_5e^{a_6\phi}\lambda_\alpha
+a_7e^{a_8\phi}(\Gamma^m)_{\alpha\beta}\psi_m^{\;\;\beta}$
and $\chi^\alpha\rightarrow a_9e^{a_{10}\phi}\chi^\alpha$, where
$a_1,\cdots,a_{10}$ are constants.}

Now we construct the effective action of D0-brane
following \cite{s3}.
When we do not take account of $\xi^I$, the action has only the
Born-Infeld part and does not have the Chern-Simons part:
\beq
S_{\rm BI}=-\frac{1}{\sqrt{2\ap}}\int d\tau \bm{\Phi}^{\frac{5}{4}}
\sqrt{\Pi_\tau^{\;\;a}\Pi_{\tau a}},
\label{biaction}
\eeq
where $\Pi_\tau^{\;\;A}=\p_\tau Z^M\bm{ E}_M^{\;\;A}$ and
$Z^M=(X^m,\theta^\alpha)$. The exponent of $\bm{\Phi}$ is determined
so that this action is proportional to $e^{-\phi}$ when we use the
string metric.

$S_{\rm BI}$ does not have $\kappa$-symmetry since it does not have
Chern-Simons term. Therefore the spacetime supersymmetry is
spontaneously broken completely. Worldsheet diffeomorphism is the only
local symmetry which $S_{\rm BI}$ has. The number of physical modes
of $\theta^\alpha$ is twice that of BPS D-branes because of the
absence of $\kappa$-symmetry.

Now we take account of $\xi^I$. In addition to $S_{\rm BI}$, we need
the following term.
\beq
S_\xi=\int d\tau\Half i\xi^I(\delta_{IJ}\p_\tau
-\p_\tau Z^M(\bm{A}_M)_{IJ})\xi^J.
\label{xiaction}
\eeq
This is determined by requiring invariance under worldsheet
diffeomorphism and $SO(32)$ gauge transformation.
The total action $S$ is the sum of $S_{\rm BI}$ and $S_\xi$.
Substituting eq.(\ref{compe1}), (\ref{compe2}), (\ref{compe3}),
(\ref{compe4}), (\ref{compa1}), (\ref{compa2}), and (\ref{compp})
into $S$ we get
\bea
S_{\rm BI} & = & -\frac{1}{\sqrt{2\ap}}\int d\tau e^{-\frac{5}{6}\phi}
\sqrt{g_{mn}\p_\tau X^m\p_\tau X^n} \nn\\
& & \times\biggl[1
+\frac{5}{4}e^{\frac{2}{3}\phi}\theta^\alpha\lambda_\alpha 
+\frac{15}{384}c_1
\theta^\alpha(\Gamma^{abc})_{\alpha\beta}\theta^\beta 
(\Gamma_{abc})_{\gamma\delta}{\rm tr}[\chi^\gamma\chi^\delta]
\nn\\ 
& & +(g_{kl}\p_\tau X^k\p_\tau X^l)^{-1}
\Bigl(-2\p_\tau X^m\p_\tau X^n
\theta^\alpha(\Gamma_m)_{\alpha\beta}\psi_n^{\;\;\beta}
-\p_\tau X^m
\theta^\alpha(\Gamma_m)_{\alpha\beta}D_\tau\theta^\beta \nn\\
& & +\frac{1}{16}g_{mn}\p_\tau X^m\p_\tau X^n
\theta^\alpha(\Gamma^{abc})_{\alpha\beta}\theta^\beta T_{abc}
+\frac{1}{4}\p_\tau X^m\p_\tau X^ne_n^{\;\; a}
\theta^\alpha(\Gamma_{mbc})_{\alpha\beta}\theta^\beta
T_a^{\;\; bc}\Bigr)\biggr] \nn\\
& & +O(\theta^3,{\rm (fermion)^2}), \\
S_\xi & = & \int d\tau\Half i\xi^I(\delta_{IJ}\p_\tau
-\p_\tau X^m(A_m)_{IJ}+\p_\tau X^m
\theta^\alpha(\Gamma_m)_{\alpha\beta}(\chi^\beta)_{IJ})\xi^J \nn\\
& & +O(\theta^2,{\rm (fermion)^2}),
\eea
where $g_{mn}=e_m^{\;\;a}e_{na}$ and
\beq
D_\tau\theta^\alpha=\p_\tau\theta^\alpha
+\frac{1}{4}\p_\tau X^m\omega_{mab}
\theta^\beta(\Gamma^{ab})_\beta^{\;\;\alpha}.
\eeq
From this action we can construct the vertex operators of
$\psi_m^{\;\;\alpha}$, $\lambda_\alpha$, $\chi^\alpha$ and
$A_m$:
\bea
\int d\tau (V_\psi)_\alpha^m\psi_m^{\;\;\alpha} & = &
\frac{1}{\sqrt{2\ap}}\int d\tau 2e^{-\frac{5}{6}\phi}
\frac{\p_\tau X^m\p_\tau X^n}{\sqrt{g_{kl}\p_\tau X^k\p_\tau X^l}}
\theta^\beta(\Gamma_m)_{\beta\alpha}\psi_m^{\;\;\alpha}
+O(\theta^2), \\
\int d\tau (V_\lambda)^\alpha\lambda_\alpha & = &
-\frac{1}{\sqrt{2\ap}}\int d\tau\frac{5}{4}e^{-\frac{1}{6}\phi}
\sqrt{g_{kl}\p_\tau X^k\p_\tau X^l}\theta^\alpha\lambda_\alpha
+O(\theta^2), \\
\int d\tau (V_\chi)^{IJ}_\alpha\chi^\alpha_{IJ} & = &
\int d\tau\Half i \xi^I\xi^J\p_\tau X^m
\theta^\beta(\Gamma_m)_{\beta\alpha}\chi^\alpha_{IJ} 
+O(\theta^2), \\
\int d\tau (V_A)^mA_m & = &
\int d\tau (-i)\Half\xi^I\xi^J\p_\tau X^mA_{mIJ}
+O(\theta^2). \\
\eea
Similarly we can construct the vertex operators of graviton, dilaton,
and RR 2-form.
These vertex operators are used for calculating instanton effects
following \cite{bbs,hm}.

\section{A simple example}

In this section we consider $S^1$ compactification and some amplitudes
as a simple example of non-BPS D0-brane instanton effects using the
prescription given in \cite{bbs,hm}. We discuss the low energy and
large radius regime in order to calculate instanton effects by the
effective action, as explained in \cite{hm}. We do not keep track of
the overall normalization
of the amplitudes since it cannot be determined by this method.

Let us take the background geometry $S^1\times R^9$
($g_{mn}$ is flat, $\phi$ is a constant and other fields are zero).  
We choose $X^9$ as the coordinate of $S^1$ and set the radius $R$. 
Substituting eq.(\ref{flate}), (\ref{flata}) and (\ref{flatp}) to
eq.(\ref{biaction}) and (\ref{xiaction}) we get the D0-brane action in
this background:
\beq
S = -\frac{1}{\sqrt{2\ap}}\int d\tau e^{-\phi}
\sqrt{(\p_\tau X^m
-\p_\tau\theta^\alpha(\Gamma^m)_{\alpha\beta}\theta^\beta)
(\p_\tau X_m
-\p_\tau\theta^\gamma(\Gamma_m)_{\gamma\delta}\theta^\delta)}
+\int d\tau\Half i\xi^I\p_\tau\xi^I.
\eeq
Here we are using the string metric.

Let us consider an Euclidean D0-brane winding around $S^1$.
If we choose the static gauge, then
\beq
X^9=R\tau,\quad 0\leq\tau\leq 2\pi.
\eeq
The action up to the quadratic order becomes
\beq
S=-\frac{2\pi R}{\sqrt{2\ap}g_s}
-\frac{1}{\sqrt{2\ap}g_sR}\int d\tau\Half(\p_\tau X^i)^2
+\frac{1}{\sqrt{2\ap}g_s}\int d\tau
\p_\tau\theta^\alpha(\Gamma^9)_{\alpha\beta}\theta^\beta
+\int d\tau\Half i\xi^I\p_\tau\xi^I,
\eeq
with $i=0,1,\cdots,8$.

The zero modes are the constant modes of $X^i$, $\theta^\alpha$ and
$\xi^I$. The zero modes of $X^i$ represent
the position of the D0-brane in $R^9$. To soak up the fermion zero
modes we must insert 16 $\theta$'s and 32 $\xi$'s.

In the flat background the vertex operators of
$\psi_9^{\;\;\alpha}(X^i)$,
$\lambda_\alpha(X^i)$, $\chi^\alpha(X^i)$ and $A_9(X^i)$ become
\bea
(V_\psi)_\alpha^9 & \sim &
 (\Gamma^9)_{\alpha\beta}\theta^\beta, \\
(V_\lambda)^\alpha & \sim & \theta^\alpha, \\
(V_\chi)^{IJ}_\alpha & \sim &
 (\Gamma_9)_{\alpha\beta}\theta^\beta\xi^I\xi^J, \\
(V_A)^{9IJ} & \sim & \xi^I\xi^J,
\eea
where we ignored the coefficients.

The simplest amplitudes saturating the fermion zero modes are as
follows. 
\begin{itemize}
\item 16 $V_\chi$ insertions \\
This gives the following correction to the effective action.
\beq
\epsilon^{\alpha_1\cdots\alpha_{16}}\epsilon^{I_1\cdots I_{32}}
(\Gamma^9)_{\alpha_1\beta_1}\chi^{\beta_1}_{I_1I_2}\cdots
(\Gamma^9)_{\alpha_{16}\beta_{16}}\chi^{\beta_{16}}_{I_{31}I_{32}}
\exp(-\frac{2\pi R}{\sqrt{2\ap}g_s})
\eeq
\item 16 $(V_A)^9$, $n$ $V_\lambda$
 and $16-n$ $(V_\psi)^9$ insertions \\
This gives the following correction to the effective action.
\bea
& &\epsilon^{\alpha_1\cdots\alpha_{16}}\epsilon^{I_1\cdots I_{32}}
(A_9)_{I_1I_2}\cdots (A_9)_{I_{31}I_{32}}
\lambda_{\alpha_1}\cdots
\lambda_{\alpha_n}
\nn\\
& & \times(\Gamma^9)_{\alpha_{n+1}\beta_{n+1}}
\psi_9^{\;\;\beta_{n+1}}\cdots 
(\Gamma^9)_{\alpha_{16}\beta_{16}}\psi_9^{\;\;\beta_{16}}
\exp(-\frac{2\pi R}{\sqrt{2\ap}g_s})
\eea
\end{itemize}

We can consider the compactification on higher dimensional torus
similarly to the above analysis. In this case the D0-brane is unstable
if the radius of the direction transverse to it is sufficiently small,
as explained in section 2. But we do not have to consider this fact
since our discussion is restricted to the large radius ($R\gg \ap$)
regime. 

\section{Discussions}

We calculated the instanton effects by the effective action.
Of course we can consider stringy calculation by using the rule given
in \cite{s2}. In this case we must be careful with the instability of
the D0-brane caused by sufficiently small radius of the direction
transverse to it.

The D0-brane corresponds to the perturbative massive modes of
heterotic $SO(32)$ theory. It is interesting to calculate the effect
corresponding to the D0-brane instanton effects in the heterotic side,
though quantitative comparison with the results in type I side is
difficult since these
are not protected from the string coupling correction.

Type I string theory has non-BPS stable D$(-1)$-brane besides D0-brane
\cite{w}.
Its instanton effects exist even in flat ten dimensional spacetime.
The properties of D$(-1)$-brane are similar to the D0-brane. It is
interesting to calculate these effects.

Instanton effects correct the low energy effective action. It may be
possible to determine the exact tensions of D0 and D$(-1)$-brane by
completing the action supersymmetrically and identifying the instanton
actions.

\vs{.5cm}
\noindent
{\large\bf Acknowledgments}\\[.2cm]
I would like to thank H.\ Hata for careful reading of the manuscript.

\newcommand{\J}[4]{{\sl #1} {\bf #2} (#3) #4}
\newcommand{\andJ}[3]{{\bf #1} (#2) #3}
\newcommand{\AP}{Ann.\ Phys.\ (N.Y.)}
\newcommand{\MPL}{Mod.\ Phys.\ Lett.}
\newcommand{\NP}{Nucl.\ Phys.}
\newcommand{\PL}{Phys.\ Lett.}
\newcommand{\PR}{Phys.\ Rev.}
\newcommand{\PRL}{Phys.\ Rev.\ Lett.}
\newcommand{\PTP}{Prog.\ Theor.\ Phys.}
\newcommand{\hepth}[1]{{\tt hep-th/#1}}

\end{document}